\documentclass[twocolumn,showpacs,preprintnumbers,amsmath,amssymb]{revtex4}

\usepackage{graphicx}
\usepackage{dcolumn}
\usepackage{bm}

\begin{document}

\preprint{APS/123-QED}

\title{Multicritical behaviour of the compressible systems}

\author{S.V. Belim}
 \email{belim@univer.omsk.su}
\affiliation{%
Omsk State University, 55-a, pr. Mira, Omsk, Russia, 644077
\textbackslash\textbackslash
}%

\date{\today}

\begin{abstract}
The behaviour of uniform elastically isotropic compressible
systems in critical and tricritical points is described in
field-theoretical terms. Renormalizationgroup equations are
analyzed for the case of three-dimensional systems in a two-loop
approximation. Fixed points
corresponding to various types of critical and multicritical
behaviour under various macroscopic conditions imposed on the
system are distinguished. It is shown that the effect of the
elastic deformations on the critical behaviour of compressible
systems is significant. It manifests itself both in a change in
the critical exponents of Ising magnetic and in the appearance of
multicritical points in phase diagrams at any dimension of the
order parameter. It is also shown that, in a number of
experimental investigations, the multicritical behaviour is not
tricritical, as it has been stated, but tetracritical. The
influence exerted by elastic deformations on systems with phase
diagrams already containing multicritical points is analyzed. It
is shown that the effect of elastic deformations leads to a change
from bicritical behaviour to a tetracritical one.

\end{abstract}

\maketitle
\section{Introduction}
Phase diagrams of many materials
exhibit multicritical behaviour when several lines of phase
transitions are crossed. The tricritical points, in which the
second-order phase transition is replaced by the first-order
transition, are of particular interest. Experimental studies are
mostly concerned with systems subjected to external pressure, i.e.
those influenced by elastic deformations. Owing to this
circumstance, taking into account the relationship between the
order parameter and elastic deformations is important. As first
shown in [1], in the elastically isotropic case, the critical
behaviour of compressible systems with quadratic striction is
unstable with respect to the relationship between the order
parameter and acoustic modes, and a first-order phase transition
close to the second-order transition occurs in systems of this
kind. However, the conclusions of [1] are only valid at
low pressures. Itwas shown in [2] that at high pressures, a more
fundamental influence is exerted on the system by deformational
effects induced by the external pressure, beginning from a certain
tricritical value $P_t$. This changes the sign of the effective
constant of interaction between fluctuations of the order
parameter, and, as a consequence, leads to a change of the type of
the phase transition. For uniform compressible systems, two types
of tricritical behaviour and the existence of a tetracritical
point, at which two tricritical curves meet, were predicted in
[2]. Calculations made in terms of a two-loop approximation [3]
confirm the existence of two types of tricritical behaviour for
the Ising systems and yield the tricritical exponents.

According to the criterion obtained in [1], striction effects, regarded as
additional thermodynamic parameters, change the mode of the
critical behaviour only in systems with a singular behaviour of
heat capacity in the absence of deformations. The specific heat
exponent $\alpha(C\sim |T -T_c|^{-/alpha})$ is only positive for Ising-like
magnetic. In the XY-model and the Heisenberg model, the specific
heat exponent is positive for "rigid" systems, and, therefore,
elastic deformations must have no effect on the critical
behaviour. Hence follows that the critical value of the order
parameter dimensionality $n_c < 2$.

In this study, we develop further
the model of phase transformations in uniform compressible systems
characterized by various dimensions of the fluctuating order
parameter [4, 5]. We analyze these transformations using the
renormalization-group methods in a two-loop approximation,
directly in the three-dimensional space. Also, we consider the
conditions for occurrence of a tricritical behaviour owing to
effects of long-range interaction of the order-parameter
fluctuations caused by long-wavelength acoustic modes. Since the
dependence of the exchange integral on distance makes a major
contribution to the striction effects in the critical region, we
consider only elastically isotropic systems.

\section{Theory}
We can write the Hamiltonian of the uniform Ising model, with allowance for elastic
deformations, as follows:
\begin{eqnarray}
&&H_0=\int d^Dx[\frac{1}{2}(\tau_1+\nabla^2)S(x)^2 +
+\frac{u_0}{4!}(S(x)^2)^2]\nonumber\\
&&+\int d^Dx[a_1(\sum\limits_{\alpha=1}^3u_{\alpha \alpha }(x))^2+a_2\sum\limits_{\alpha ,\beta
=1}^3u_{\alpha \beta }^2]+\nonumber\\
 &+&\frac 12a_3\int d^Dx\vec{S}(x)^2 (\sum\limits_{\alpha =1}^3u_{\alpha\alpha }(x))
\end{eqnarray}
where $S(x)$ is the $n$-dimensional order parameter; $u_0$ is a positive constant;
$\tau_0\sim|T-T_c|/T_c$, where $T_c$ is the phase transition temperature;
$u_{\alpha \beta}$ is the deformation tensor; $a_1, a_2$ are the elastic
constants of the crystal; and $a_3$ is the quadratic-striction parameter.
Passing in equation (1) to Fourier transforms and integrating with respect
to terms depending on non-fluctuating variables, which do not interact with
the parameter of the $S(x)$ order, and introducing for
the sake of convenience a new variable
$y(x)=\sum\limits_{\alpha=1}^3u_{\alpha\alpha}(x)$, we obtain the Hamiltonian
of the system in the following form:
\begin{eqnarray}
&&H_0=\frac 12\int d^Dq(\tau _0+q^2)S_qS_{-q}\nonumber\\
&&+\frac{u_0}{4!}\int d^D{q_i}S_{q1}S_{q2}S_{q3}S_{-q1-q2-q3}+\nonumber\\
&&+a_3\int d^Dqy_{q1}S_{q2}S_{-q1-q2}\\
&&+\frac{a_3^{(0)}}\Omega y_0\int d^DqS_qS_{-q}+\frac 12a_1
\int d^Dqy_qy_{-q} +\frac 12\frac{a_1^{(0)}}\Omega y_0^2\nonumber
\end{eqnarray}
In equation (2) the terms y0, which describe uniform deformations, are separated. As shown
in [1], such a separation is necessary, since the nonuniform deformations $y_q$ are responsible
for acoustic phonon exchange and lead to long-range effects, which are absent in the case of
uniform deformations.

Let us define the effective Hamiltonian of the system, which depends only on the strongly
fluctuating order parameter $S$, in the following way:
\begin{eqnarray}\label{3}
\exp \{-H[S]\}=B\int \exp \{-H_{R}[S,y]\}\prod dy_q
\end{eqnarray}
If an experiment is carried out at constant volume, $y_0$ is a constant and the integration in
equation (3) is only done over nonuniform deformations, with uniform deformations making
no contribution to the effective Hamiltonian. At a constant pressure, a term $P\Omega$
is added to the Hamiltonian, with the volume represented in terms of deformation tensor
components as
\begin{eqnarray}\label{6_1}
\Omega=\Omega_0 [1+\sum\limits_{\alpha =1}u_{\alpha\alpha}+
\sum\limits_{\alpha \neq \beta}u_{\alpha\alpha}u_{\beta\beta}+O(u^3)]
\end{eqnarray}
and integration in equaation (3) also performed over uniform deformations. As noted in [6],
taking into account the quadratic terms in equation (4) can be important in the case of high
pressures and crystals with large striction effects. The neglect of these quadratic terms in [1]
restricts applicability of the results obtained by Larkin and Pikin to only the case of low
pressures. As a result, we have
\begin{eqnarray}\label{7}
&&H=\frac 12\int d^Dq(\tau _0+q^2)S_qS_{-q}\\
&&+\Big(\frac{u_0}{4!}- \frac{z_0}{2}\Big)\int d^D\{q_i\}S_{q1}S_{q2}
S_{q3}S_{-q1-q2-q3}\nonumber\\
&+&\frac{1}{2\Omega}(z_0 - w_0)\int d^D\{q_i\}S_{q1}S
_{-q1}S_{q2}S_{-q2},\nonumber\\
&&z_0 =a_1^2/(4a_3),  \ \  w_0 =a_1^{(0)2}/(4a_3^{(0)}). \nonumber
\end{eqnarray}
The effective interaction parameter $v_0=u_0-12z_0$, which appears in the Hamiltonian owing
to the influence of striction effects described by the parameter $z_0$,
can assume not only positive, but also negative values. As a result, this Hamiltonian
describes both first- and second-order phase transitions. At $v_0=0$, a tricritical
behaviour is exhibited by the system. In its turn, the effective interaction in equation (5),
which is determined by the difference of the parameters $z_0 - w_0$, may lead to
a second-order phase transition at $z_0 - w_0 > 0$ and to a first-order
phase transition at $z_0 -w_0 < 0$. From the given type of the effective Hamiltonian
follows that there can exist a higher-order critical point at which tricritical
curves intersect if the conditions $v_0=0$ and $z_0=w_0$ are fulfilled simultaneously [2].
It should be noted that, under the tricritical condition $z_0=w_0$, the Hamiltonian
of the model (5) is isomorphic to the Hamiltonian of a uniform rigid system.

In terms of the field-theoretical approach [7], the asymptotic critical behaviour and the
structure of phase diagrams in the fluctuation region are determined by the Callan–Symanzyk
renormalization-group equation for the vertex parts of irreducible Green function. In order
to calculate $\beta$- and $\gamma$-functions as functions appearing in the Callan–Symanzyk
equation of renormalized interaction vertices $u, a_1, a_1^{(0)}$ or as functions of complex
vertices $z =a_1^{2}/(4a_3)$, $w =a_1^{(0)2}/(4a_3^{(0)})$, $v=u-12z$, which are more
convenient for determining the critical and tricritical behaviour of the model,
we applied the standard method based on Feynman's diagram technique and the renormalization
procedure [8]. As a result, we obtained in terms of the two-loop approximation the following
expressions for the $\beta$-functions
\begin{eqnarray}\label{8}
\beta _v&=&-v\Big{(} 1-\frac{n+8}{6}v+\frac{41n+190}{243}v^2\Big{)},\nonumber\\
\beta _z &=&-z\Big{(}1-\frac{n+2}{3}v-2nz+\frac{23(n+2)}{243}v^2\Big{)},\\
\beta _w &=&-w\Big{(}1-\frac{n+2}{3}v-4nz+2nw+\frac{23(n+2)}{243}v^2\Big{)}.\nonumber
\end{eqnarray}
It is known that the perturbation-theory expansions are asymptotic and the interaction
vertices of order-parameter fluctuations in the fluctuation region are sufficiently large
for equations (14) to be applied directly. Therefore, in order to extract the necessary
physical information from the expressions obtained, we use the Pad´e–Borel method extended
to a three-parameter case. Then the direct and inverse Borel transformations have the
following form:
\begin{equation}
\begin{array}{rl} \displaystyle
  & f(v,z,w)=\sum\limits_{i_1,i_2,i_3}c_{i_1,i_2,i_3}v^{i_1}z^{i_2}w^{i_3}\nonumber\\
  &=\int\limits_{0}^{\infty}e^{-t}F(vt,zt,wt)dt,  \\
  & F(v,z,w)=\sum\limits_{i_1,i_2,i_3}\frac{\displaystyle c_{i_1,i_2,i_3}}{\displaystyle(i_1+i_2+i_3)!}v^{i_1}z^{i_2}w^{i_3}.
\end{array}
\end{equation}
To obtain an analytical continuation of the Borel transform of the function, we introduce a
series in auxiliary variable $\theta$:
\begin{equation}
   {\tilde{F}}(v,z,w,\theta)=\sum\limits_{k=0}^{\infty}\theta^k\sum\limits_{i_1,i_2,i_3}\frac{\displaystyle c_{i_1,i_2,i_3}}{\displaystyle k!}v^{i_1}z^{i_2}w^{i_3}\delta_{i_1+i_2+i_3,k}\  ,
\end{equation}
to which we apply the Pade [L/M] approximation at the point $\theta=1$. This technique was
proposed and tested in [9] for describing the critical behaviour of a number of systems
characterized by several interaction vertices of order-parameter fluctuations. The property
of symmetry conservation of the system upon application of the Pade approximant in variable
$\theta$, revealed in [9], becomes significant in describing multivertex models.

In the two-loop approximation, we used the [2/1] approximant for calculating the
$\beta$-functions. The nature of the critical behaviour is determined by the existence
of a stable fixed point that satisfies the set of equations
\begin{equation}
\beta_{i}(v^*,z^*,w^*)=0\ \ \ \ \ \ \ \ \ \ \ \ \ \   (i=1,2,3).
\end{equation}
The requirement that a fixed point be stable is reduced to the condition that
the eigenvalues $b_i$ of the matrix
\begin{equation}
B_{i,j}=\frac{\partial\beta_i(u_1^*,u_2^*,u_3^*)}{\partial{u_j}}\ \ \ \ \ \ \ \ \  (u_i,u_j \equiv v,z,w)
\end{equation}
lie in the right-hand complex half-plane. The fixed point with $v^*=0$, corresponding to
the critical behaviour, is a saddle point and must be stable in the directions specified
by the variables $z$, $w$ and unstable in the direction specified by the variable $v$.
The tricritical fixed point is stabilized in the direction specified by the variable $v$
as a result of taking into account in the effective Hamiltonian of the model terms
of sixth order in order-parameter fluctuations. The fixed point with $z^*=w^*$,
which corresponds to the tricritical behaviour of the second type, is also a saddle
point and must be stable in the directions specified by the variables $v, z$
and unstable in the direction specified by the variable $w$. Its stabilization may be due to
anharmonic effects.

The obtained set of the summed-up functions contains a wide variety of fixed points.
Table 1 presents fixed points for the Ising model ($n=1$), XY model ($n=2$) and the
Heisenberg model ($n=3$), which are the most interesting for describing the critical
and tricritical types of behaviour, which lie in the physical range of vertices
with $v,  z, w \geq 0$. The table also contains the eigenvalues of the stability matrix
for the corresponding fixed points.

\begin{table*}
\hspace{120mm}
\vspace{5mm}
\begin{center}
Table 1. Magnitudes of fixed points and eigenvalues of the stability matrix.
\end{center}
\begin{center}
\begin{tabular}{|c|c|c|c|c|c|c|c|c|} \hline
 N & $v^{*}$ &$z^{*}$    &$w^{*}$    &$b_{1}$    &$b_{2}$     &$b_{3}$     \\
\hline
\multicolumn{7}{|c|} {n=1} \\
\hline
 I0 & 0        &  0        &      0    & -1        & -1         & -1         \\
 I1 & 1.064472 &  0        &      0    &  0.6536   & -0.1692    &  -0.1692   \\
 I2 & 1.064472 &  0.089187 &      0    &  0.6536   & 0.1702     &  0.1710    \\
 I3 & 1.064472 &  0.089187 & 0.089187  &  0.6536   & 0.1702     &  -0.1710   \\
 I4 & 0        &  0.5      &      0    &  -1       & 1          &  -0.16923  \\
 I5 & 0        &  0.5      &      0.5  &  -1       & 1          &  -1        \\
\hline
\multicolumn{7}{|c|} {n=2}\\
\hline
 X0 & 0        &  0        &      0    & -1        & -1         & -1         \\
 X1 & 0.934982 &  0 & 0  &  0.6673  & -0.0017   &  0.1053 \\
 X2 & 0.934982 &  0.000439 & 0   & 0.6673  & 0.0017 & 0.1087 \\
 X3 & 0.934982 & 0.000439 & 0.000439& 0.6673   &0.0017 & -0.1053\\
 X4 & 0  &  0.25 & 0 &-1 & 1& 1   \\
 X5 & 0  &  0.25 & 0.25 &-1&1&-1 \\
 \hline
\multicolumn{7}{|c|} {n=3}\\
\hline
 G0 & 0        &  0        &      0    & -1     & -1      & -1     \\
 G1 & 0.829620 &  0        &      0    & 0.6813 & 0.1315  & 0.2173 \\
 G2 & 0.829620 &  0.022909 &      0    & 0.6813 & -0.1311 & -0.0518 \\
 G3 & 0.829620 &  0.022909 & 0.022909  & 0.6813 & -0.1311 & -0.2170 \\
 G4 & 0        &  1/6      &      0    & -1     & 1       & 1      \\
 G5 & 0        &  1/6      &     1/6   & -1     & 1       & -1      \\
\hline
\end{tabular} \end{center} \end{table*}

Analysis of the magnitudes of fixed points and their stability suggests the following:
the Gaussian fixed points I0, X0, G0 are tricritical and unstable with respect to the
effect of elastic deformations. The critical behaviour of incompressible systems with
respect to the deformation degrees of freedom is unstable for the Ising model (I1)
and stable for the Heisenberg model (G1). For theXYmodel (X1), the eigenvalue $b_2 < 0$,
but is comparable in order of magnitude with the accuracy of calculations;
therefore, we cannot make any unambiguous conclusion concerning
the stability of this fixed point. Apparently, the difficulties in description of the
XY model are associated with the closeness of the critical dimensionality of the order
parameter $n_c$ to 2. According to the criterion obtained in [1], $n_c < 2$,
whereas the two-loop approximation gives $n_c=2.011$. For Ising systems, the fixed
point is stable at constant pressure (I2); for Heisenberg systems, the corresponding point
is unstable (G2); for the XY model, it is impossible to make any unambiguous conclusion
since the critical dimensionality is also close to 2. The fixed points I3, X3 and G3
describe the first type of the tricritical behaviour of compressible systems,
which can be observed at constant pressure. The fixed points I4, X4
and G4 are tricritical for systems investigated at constant volume.
The points I5, X5 and G5 are critical points of fourth order, with two critical
lines intersecting at these points.

The magnitudes of vertices, obtained in the two-loop approximation for the fixed points
corresponding to the critical and tricritical types of behaviour of the compressible Ising
model, make it possible to calculate the critical exponent for the given systems
on the basis of the expressions (summed by the Pade–Borel method) for the exponents
$\nu$ and $\eta$:
\begin{eqnarray}
&&\nu =\frac 12\Big(1+\frac{n+2}{12}v^*+nz^* -nw^*-\frac{11(n+2)}{3888}v^{*2}\Big),\nonumber\\
&&\eta =\frac{2(n+2)}{243}v^{*2}.
\end{eqnarray}
The values of the other critical exponents can be obtained from scaling expressions relating
them to the exponents $\nu$ and $\eta$.

The critical behaviour of compressible Ising systems at constant pressure (12) is
characterized by renormalized critical exponents according to Fisher's theory of the effect
of additional thermodynamic variables [10]
\begin{eqnarray}
\nu^{(I)}=0.632,\eta^{(I)}=0.028,\alpha^{(I)}=0.103,\nonumber\\
\beta^{(I)}=0.325,\gamma^{(I)}=1.247.\nonumber
\end{eqnarray}
For the critical behaviour of the first type (I3, X3, G3) the Hamiltonian (5) is isomorphic to
the Hamiltonian of the uniform incompressible model, and, therefore, the critical exponents
coincide in this case with those for the incompressible model:
\begin{eqnarray}
&&\nu^{(I)}=0.708,\eta^{(I)}=0.028,\alpha^{(I)}=-0.125,\nonumber\\
&&\beta^{(I)}=0.364,\gamma^{(I)}=1.397,\nonumber\\
&&\nu^{(XY)}=1,\eta^{(XY)}=0,\alpha^{(XY)}=-1,\nonumber\\
&&\beta^{(XY)}=0.5,\gamma^{(XY)}=2,\nonumber\\
&&\nu^{(G)}=1,\eta^{(G)}=0,\alpha^{(G)}=-1,\nonumber\\
&&\beta^{(G)}=0.5,\gamma^{(G)}=2.\nonumber
\end{eqnarray}
The tricritical behaviour of the second type (I4, X4, G4) corresponds to the critical behaviour
of the spherical model and is determined by the corresponding exponents:
\begin{eqnarray}
\nu^{(I)}=1,\eta^{(I)}=0,\alpha^{(I)}=-1,\nonumber\\
\beta^{(I)}=0.5,\gamma^{(I)}=2,\nonumber\\
\nu^{(XY)}=1,\eta^{(XY)}=0,\alpha^{(XY)}=-1,\nonumber\\
\beta^{(XY)}=0.5,\gamma^{(XY)}=2,\nonumber\\
\nu^{(G)}=1,\eta^{(G)}=0,\alpha^{(G)}=-1,\nonumber\\
\beta^{(G)}=0.5,\gamma^{(G)}=2.\nonumber
\end{eqnarray}
The fixed points of the fourth order (I4, X4, G4) are characterized by the mean-field values of
the critical exponents:
\begin{eqnarray}
\nu^{(I)}=0.5,\eta^{(I)}=0,\alpha^{(I)}=0.5,\nonumber\\
\beta^{(I)}=0.25,\gamma^{(I)}=1,\nonumber\\
\nu^{(XY)}=0.5,\eta^{(XY)}=0,\alpha^{(XY)}=0.5,\nonumber\\
\beta^{(XY)}=0.25,\gamma^{(XY)}=1,\nonumber\\
\nu^{(G)}=0.5,\eta^{(G)}=0,\alpha^{(G)}=0.5,\nonumber\\
\beta^{(G)}=0.25,\gamma^{(G)}=1.\nonumber
\end{eqnarray}
Thus, the system may exhibit a multicritical behaviour under the influence of striction effects.
This poses the question as to how elastic deformations affect systems whose phase diagrams
already contain multicritical points of bi- or tetracritical nature. Two lines of second-order
phase transitions and a line of a first-order phase transition intersect at the multicritical point
in the first case, and four lines of second-order phase transitions, in the second. In the
immediate vicinity of the multicritical point, the system manifests a specific critical behaviour
characterized by the competition of ordering types. In this case, one critical parameter is
replaced by another at the bicritical point, whereas the tetracritical point allows existence of
a mixed phase with coexisting types of ordering. Systems of this kind [13] can be described
by introducing two order parameters transformed in accordance with different irreducible
representations.

In this case, the model Hamiltonian of the system is as follows:
\begin{eqnarray}\label{gam1}
&&H_0=\int d^Dx\Big[\frac{1}{2}(\tau_1+\nabla^2)\Phi(x)^2
+\frac{1}{2}(\tau_1+\nabla^2)\Psi(x)^2\nonumber\\
&&+\frac{u_{10}}{4!}(\Phi(x)^2)^2
+\frac{u_{20}}{4!}(\Psi(x)^2)^2
+\frac{2u_{30}}{4!}(\Phi(x)\Psi(x))^2\nonumber\\
&&+g_1y(x)\Phi(x)^2
+g_2y(x)\Psi(x)^2
+\beta y(x)^2\Big],
\end{eqnarray}
where $\Phi(x)$ and $\Psi(x)$ are fluctuating order parameters;
$u_{10}$ and $u_{20}$ are positive constants;
$\tau_1\sim|T-T_{c1}|/T_{c1}$, $\tau_2\sim|T-T_{c2}|/T_{c2}$,
where $T_{c1}$ and $T_{c2}$ are the phase-transition temperatures
for, respectively, the first and second order parameters;
$y(x)=\sum\limits_{\alpha=1}^3u_{\alpha\alpha}(x)$,
where $u_{\alpha\beta}$ is the deformation tensor;
$g_1$ and $g_2$ are the quadratic-striction parameters; $\beta$ is a constant
characterizing the elastic properties of the crystal; and $D$ is the space
dimensionality. As shown above, the striction effects are significant only
for Ising systems. Therefore, we consider here only the case of one-dimensional
order parameters.

Performing transformations analogous to the case of a single order parameter, we come
to the following effective Hamiltonian:
\begin{eqnarray}\label{gam3}
&&H =\frac 12\int d^Dq(\tau _1+q^2)\Phi_q\Phi_{-q}\nonumber\\
&&+\frac 12\int d^Dq(\tau _2+q^2)\Psi_q\Psi_{-q}\\
&&+\frac{v_{01}}{4!}\int d^D{q_i}(\Phi_{q1}\Phi_{q2})(\Phi_{q3}\Phi_{-q1-q2-q3})\nonumber\\
&&+\frac{v_{02}}{4!}\int d^D{q_i}(\Psi_{q1}\Psi_{q2})(\Psi_{q3}\Psi_{-q1-q2-q3})\nonumber\\
&&+\frac{2v_{03}}{4!}\int d^D{q_i}(\Phi_{q1}\Phi_{q2})(\Psi_{q3}\Psi_{-q1-q2-q3})\nonumber\\
&&+\frac{z_1^2-w_1^2}{2}\int d^D{q_i}(\Phi_{q1}\Phi_{-q1})(\Phi_{q2}\Phi_{-q2})\nonumber\\
&&+\frac{z_2^2-w_2^2}{2}\int d^D{q_i}(\Psi_{q1}\Psi_{-q1})(\Psi_{q2}\Psi_{-q2})\nonumber\\
&&+(z_1z_2-w_1w_2)\int d^D{q_i}(\Phi_{q1}\Phi_{-q1})(\Psi_{q2}\Psi_{-q2})\nonumber \\
&& v_{01}=u_{01}-12z_1^2, \ \ v_{02}=u_{02}-12z_2^2, \nonumber\\
&& v_{03}=u_{03}-12z_1z_2,\nonumber \\
&& z_1=\frac{g_1}{\sqrt\beta},\ \ z_2=\frac{g_2}{\sqrt\beta},\ \
  w_1=\frac{g^0_1}{\sqrt\beta_0},\ \ w_2=\frac{g^0_2}{\sqrt\beta_0}\nonumber
\end{eqnarray}
This Hamiltonian leads to a wide variety of multicritical points. As in the case of
incompressible systems, both tetracritical $(v_3+12(z_1z_2-w_1w_2))^2<(v_1+12(z_1^2-w_1^2))(v_2+12(z_2^2-w_2^2))$
and bicritical $(v_3+12(z_1z_2-w_1w_2))^2\geq(v_1+12(z_1^2-w_1^2))(v_2+12(z_2^2-w_2^2))$
types of behaviour are possible. In addition, the striction effects may lead to
multicritical points of higher order.

In order to calculate functions as functions appearing in the Callan–Symanzyk equation
of renormalized interaction vertices $u_1, u_2, u_3, g_1, g_2, g_1^{(0)}, g_2^{(0)}$
or as functions of complex vertices $z_1$, $z_2$,  $w_1$, $w_2$, $v_1$, $v_2$, $v_3$,
which are more convenient for determining the multicritical
behaviour of the model, weapplied the standard method based on Feynman's diagram technique
and on a renormalization procedure [8]. As a result, we obtained in terms of the two-loop
approximation the following expressions for the functions:
\begin{eqnarray}\label{8}
&&\beta _{v1}=-v_1+\frac{3}{2}v_1^2+\frac{1}{6}v_3^2-\frac{77}{81}v_1^3
-\frac{23}{243}v_1v_3^2-\frac{2}{27}v_3^3,\nonumber\\
&&\beta _{v2}=-v_2+\frac{3}{2}v_2^2+\frac{1}{6}v_3^2-\frac{77}{81}v_2^3
-\frac{23}{243}v_2v_3^2-\frac{2}{27}v_3^3,\nonumber\\
&&\beta _{v3}=-v_1+\frac{2}{3}v_3^2+\frac{1}{2}v_1v_3+\frac{1}{2}v_2v_3\\
&&-\frac{41}{243}v_3^3-\frac{23}{162}v_1^2v_3
-\frac{23}{162}v_2^2v_3-\frac{1}{3}v_1v_3^2-\frac{1}{3}v_2v_3^2,\nonumber\\
&&\beta _{z1}=-z_1+v_1z_1+2z_1^3+2z_1z_2^2+\frac{1}{3}v_3z_2
-\frac{23}{81}v_1^2z_1\nonumber\\
&&-\frac{7}{243}v_3^2z_1-\frac{2}{27}v_3^2z_2,\nonumber\\
&&\beta _{z2}=-z_2+v_2z_2+2z_2^3+2z_1^2z_2+\frac{1}{3}v_3z_1
-\frac{23}{81}v_2^2z_2\nonumber\\
&&-\frac{7}{243}v_3^2z_2-\frac{2}{27}v_3^2z_1,\nonumber\\
&&\beta _{w1}=-w_1+\frac{1}{3}v_1w_1+4z_1^2w_1-2w_1^3+4z_1z_2w_2\nonumber\\
&&-2w_1w_2^2+\frac{1}{3}v_3w_2-\frac{23}{81}v_1^2w_1-\frac{7}{243}v_3^2w_1-\frac{2}{27}v_3^2w_2,\nonumber\\
&&\beta _{w2}=-w_2+v_2w_2+4z_2^2w_2-2w_2^3+4z_1z_2w_1\nonumber\\
&&-2w_1^2w_2+\frac{1}{3}v_3w_1-\frac{23}{81}v_2^2w_2-\frac{7}{243}v_3^2w_2-\frac{2}{27}v_3^2w_1.\nonumber
\end{eqnarray}
The obtained set of summed-up functions contains a wide variety of fixed points, which
lie in the physical range of vertices with $v_i\geq 0$.

Analysis of the magnitudes of fixed points and of their stability suggests the following.
The bicritical fixed point of incompressible systems ($v_1=0.934982$, $v_2=0.934982$,
$v_3=0.934982$, $z_1=0$, $z_2=0$, $w_1=0$, $w_2=0$)
is unstable with respect to the effect of uniform deformations
($b_1=0.090$, $b_2=0.523$, $b_3=0.667$,
$b_4=-0.521$, $b_5=-0.002$, $b_6=-0.521$, $b_7=-0.002$).
Striction effects stabilize the tetracritical fixed point for compressible systems
($v_1=0.934982$, $v_2=0.934982$, $v_3=0.934982$,
$z_1=0$, $z_2=0$, $w_1=0$, $w_2=0$, $b_1=0.090$, $b_2=0.523$, $b_3=0.667$,
$b_4=2.144$, $b_5=0.267$, $b_6=5.223$, $b_7=0.882$).

The question of whether or not other multicritical points exist cannot be resolved in terms
of the model described here since the calculations lead to a degenerate system of equations.
The degeneration is lifted when terms of higher order in deformation tensor components and
in fluctuating order parameters are taken into account in the Hamiltonian.

The investigation we performed revealed a significant influence of elastic deformations
on the critical behaviour of compressible systems, which is manifested both in a change in the
magnitudes of critical exponents for the Ising systems and in the appearance of multicritical
points in the phase diagrams for all the three models. The boundary value of the order parameter
dimensionality is close to 2, as in the case of the influence of frozen defects. This conclusion is
in agreement with the criterion obtained in [1] on the basis of the mean-field theory. The critical
exponents for compressible systems are in good agreement with Fisher's theory [10] of the
influence exerted by additional thermodynamic variables. All the experimental studies of the
multicritical behaviour we are aware of have been carried out for compressible systems [11].
All of them lead to a conclusion that the critical behaviour is characterized by a Gaussian
fixed point with mean-field magnitudes of the critical exponents. In a number of papers, an
opinion has been expressed that the system demonstrates a tricritical behaviour [12], while
in others, it has been stated that this behaviour is bicritical. The critical exponents found in
this study led us to conclude that a tetracritical behaviour determined by a fourth-order critical
point was observed in the studies mentioned above. In systems, described by two fluctuating
order parameters, the striction interaction with elastic deformations leads to a change of the
bicritical behaviour for a tetracritical one.

The work is supported by Russian Foundation for Basic Research
N 04-02-16002.

\newpage
\def\baselinestretch{1.0}

\end{document}